\documentclass{article}
\usepackage{ijcai19}

\usepackage{times}
\usepackage{soul}
\usepackage{url}
\usepackage[hidelinks]{hyperref}
\usepackage[utf8]{inputenc}
\usepackage[small]{caption}
\usepackage{graphicx}
\usepackage{amsmath}
\usepackage{booktabs}
\relax
\usepackage{times}  
\usepackage{helvet}  
\usepackage{courier}  
\usepackage{url}  
\usepackage{graphicx}  

\usepackage{xcolor}
\usepackage{soul}
\usepackage[utf8]{inputenc}
\usepackage{caption}
\usepackage[ruled,vlined,linesnumbered]{algorithm2e}
\usepackage{algpseudocode}
\usepackage{amsmath}
\usepackage{graphics}
\usepackage{epsfig}
\usepackage{multirow}
\usepackage{booktabs}
\usepackage{float}
\usepackage{latexsym}
\usepackage{subfigure}
\usepackage{longtable,pdflscape}
\usepackage{epstopdf}
\usepackage{amsthm}
\usepackage{algorithmicx}
\usepackage{algpseudocode} 
\usepackage{multirow}
\usepackage[misc]{ifsym}
\usepackage{amsmath}
\usepackage{graphics}
\usepackage{epsfig}
\usepackage{multirow}
\usepackage{booktabs}
\usepackage{float}
\usepackage{latexsym}
\usepackage{amsmath}
\usepackage[export]{adjustbox}
\usepackage{multirow}
\usepackage{booktabs}
\usepackage{array}
\newcommand{\PreserveBackslash}[1]{\let\temp=\\#1\let\\=\temp}
\newcolumntype{C}[1]{>{\PreserveBackslash\centering}p{#1}}
\newcolumntype{R}[1]{>{\PreserveBackslash\raggedleft}p{#1}}
\newcolumntype{L}[1]{>{\PreserveBackslash\raggedright}p{#1}}
\usepackage{amssymb}
\usepackage{graphicx}
\usepackage[ruled,linesnumbered]{algorithm2e}
\usepackage{setspace}
\usepackage[misc]{ifsym}

\begin{document}

\title{An Exact Algorithm for Minimum Weight Vertex Cover Problem in Large Graphs}
\author{
Luzhi Wang$^1$
\and
Chu-Min Li$^2$\and
Junping Zhou$^{*1}$\and
Bo Jin$^3$\And
Minghao Yin$^1$
\affiliations
$^1$Northeast Normal University, Changchun,China\\
$^2$MIS, Université de Picardie Jules Verne, France\\
$^3$Dalian University of Technology, Dalian, China\\
\emails
\{wanglz846,zhoujp877,ymh\}@nenu.edu.cn,
chu-min.li@u-picardie.fr,
jinbo@dlut.edu.cn
}

\maketitle
\begin{abstract}
\begin{quote}
This paper proposes a novel branch-and-bound (BMWVC) algorithm to exactly solve the minimum weight vertex cover problem (MWVC) in large graphs. The original contribution is several new graph reduction rules, allowing to reduce a graph $G$ and the time needed to find a minimum weight vertex cover in $G$. Experiments on large graphs from real-world applications show that the reduction rules are effective and the resulting BMWVC algorithm outperforms relevant exact and heuristic MWVC algorithms.
 
\end{quote}
\end{abstract}

\section{Introduction}
Nowadays, very large graphs are often used in information technology to model things such as computer networks, social networks, mobile call networks, and biological networks. These graphs usually share common characteristics: very low density, a huge number of vertices, and the vertex degrees following some statistical property \cite{newman2003structure}. Recently, algorithms in large sparse graphs have become an extensively studied topic with a large number of applications. Vertex cover problem is one of the earliest and most common graph problems. A \textit{vertex cover} of an undirected graph $G = (V, E)$, where $V$ is a set of vertices and $E$ is a set of edges, is a subset of vertices $S \subseteq V$ such that every edge of $E$ has at least one of its endpoint vertices in $S$. The \textit{minimum vertex cover problem} (MVC) of $G$ is to find a vertex cover with the smallest number of vertices. An important generalized version of MVC is the \textit{minimum weight vertex cover problem} (MWVC), in which a weight is associated with each vertex of $G$, and the problem is to find a vertex cover of minimum total weight in $G$.

MWVC is NP-hard \cite{garey1979guide}, and is closely related to the well-known maximum weight independent set problem (MWIS) and the maximum weight clique problem (MWC). MWVC, MWIS and MWC are three important vertex-weighted graph problems that play a vital role in numerous real-world applications such as wireless communication \cite{wang2015ptas,wang2017ptas}, 
network flows \cite{benlic2013breakout}, protein structure prediction \cite{mascia2010predicting}, coding theory \cite{zhian2013increasing}, and computer vision \cite{ma2013graph}.

A remarkable number of local search algorithms for MWVC have been proposed in the literature, including the ant colony optimization algorithms \cite{shyu2004ant,jovanovic2011ant},
simulated annealing algorithm \cite{xu2006efficient}, randomized gravitational emulation search algorithm \cite{balachandar2009meta,bouamama2012population}, tabu search algorithm \cite{Zhou2016Multi}, asymmetric game algorithm \cite{tang2017asymmetric}, support ratio algorithm \cite{balaji2010effective}, message passing algorithm \cite{nakajima2018towards}, local search based algorithms DLSWCC \cite{li2016efficient}, NuMWVC \cite{li2018numwvc}, Dynwvc1 and Dynwvc2 \cite{cai2018improving}, etc.

Although local search algorithms can find solutions of good quality, exact algorithms are needed to guarantee the optimality of the solutions. Many existing exact algorithms for MVC and MWVC are FPT (fixed-parameter tractable) algorithms, such as \cite{niedermeier2003efficient},\cite{Kratsch2009Fixed}, and \cite{iwata2014linear}... Although these algorithms are well studied  theoretically, they are not really used in practice to solve MVC and MWVC. As far as we know, the only one exact solver specially designed for really solving MVC is B$\&$R \cite{akiba2016branch}, and no exact algorithm is specially designed to solve MWVC in practice. Note that MWVC is much more complicated than MVC, and effective techniques for MVC implemented in B$\&$R are not applicable or ineffective for MWVC because of vertex weights.

Since there are a number of efficient exact algorithms for MWC, see e.g., \cite{jiang2017exact,jiang2018two,hebrard2018conflict}, an approach to solve MWVC exactly in a small or medium graph $G$ is to find a maximum clique $C$ in the complementary graph of $G$ and deduce a minimum vertex cover of $G$ from $C$. However, this approach is not feasible for a large sparse graph, because the complementary graph is very dense and the current exact MWC algorithms are not able to solve MWC in large dense graphs. Another approach to solve MWVC exactly in a small or medium graph $G$ is to encode it into a series of SAT instances and then use a SAT solver to find a minimum weight vertex cover \cite{xu2016new}.

In this paper, we propose an exact solver to solve MWVC in large graphs. The proposed algorithm is called BMWVC because it is based on the branch-and-bound scheme.  BMWVC has three main components. The first component uses 4 novel reduction rules to simplify a large vertex weighted graph. The second component is a heuristic to select the branching vertex. The third component is a lower bound to prune non-promising branches, by computing the weights of disjoint cliques in the graph. To the best of our knowledge, BMWVC is the first exact solver specially designed to solve MWVC exactly in large real-world graphs without calling a solver for another problem such as SAT or MWC.

BMWVC is evaluated on real-world graphs from Network Data Repository and benchmarks designed for practical applications of MWC. The results show that the reduction rules we proposed are highly effective and BMWVC greatly outperforms relevant
exact and heuristic algorithms on large graphs.

\section{Preliminaries}
Let $G = (V, E, w)$ be a vertex-weighted undirected graph with a set $V$ of vertices, a set $E$ of edges, and a weight function $w$ that assigns a non-negative integer called \textit{weight} to each vertex. The weight of a vertex $v$ is represented by $w(v)$. The weight of a vertex cover $S$, denoted by $w(S)$, is the total weights of the vertices in $S$. The \textit{minimum weight vertex cover problem} (MWVC) consists in finding a vertex cover of minimum weight in $G$, denoted by $S_m$.

The \textit{neighbourhood} of a vertex $v$ in $G$, denoted by $N(v)$, is the set containing all vertices adjacent to $v$, i.e., $N(v)=\{u \in V : \{u, v\}\in E\}$. The \textit{closed neighbourhood} of $v$, denoted by $N^*(v)$, is the set $N^*(v)=N(v) \cup \{v\}$.  The degree of $v$ is defined as the cardinality of $N(v)$, i.e., $d(v)=| N(v) |$. Let $i$ be an integer, $N_i(v)$ denotes the subset of the vertices in $N(v)$ with degree $i$, i.e., $N_i(v)=\{u|(u\in N(v))\wedge\ (d(u)=i)\}$. $N^*_i(v)$ denotes $N_i(v) \cup \{v\}$.

Let $T$ be a subset of $V$. $G[T]$ denotes the subgraph of $G$ induced by $T$. The neighbourhood of $T$, denoted by $N(T)$ is defined to be $\{u|(u\in N(v))\wedge\ (v\in T)\}$. We use $w(T)$ to denote the total weight of the vertices in $T$ and $G\setminus T$ to denote $G[V\setminus T]$. $G\setminus T$ is in fact obtained from $G$ by removing all vertices of $T$ and all edges with at least a vertex in $T$. We say that $T$ covers an edge $(u, v)$ if $u\in T$ or $v\in T$. 

An independent set $I$ is a subset of vertices of $G$ among which there are no edges. A clique $C$ is a subset of vertices of $G$ in which every two vertices are adjacent. The complement graph of $G$ is $\overline G=(V, \overline E)$, where $\overline E = \{(u, v) | u, v \in V\wedge u \neq v\wedge(u, v) \notin E\}$. A clique of $G$ is an independent set of $\overline G$ and vice versa. If $S$ is a vertex cover of $G$, then there are no edges among the vertices in $V\setminus S$. So, $V\setminus S$ is an independent set of $G$. For a minimum weight vertex cover $S_m$ of $G$, $V\setminus S_m$ is a maximum weight independent set of $G$ or a maximum weight clique of $\overline G$.

The density $D$ of $G$ is computed as $2\times |E|/(|V|\times (|V|-1))$ and the density of $\overline G$ is equal to $1-D$. It is clear that searching for a minimum weight vertex cover in $G$ could be implemented by searching for a maximum weight clique in $\overline G$. Unfortunately, when $G$
is large and sparse, searching for a maximum weight clique in $\overline G$ is very hard in practice, because $\overline G$ is large and dense. For example, if $|V| = 10^6$ and $|E|=10^{10}$ ($D \simeq 0.02$), then $\overline G$ contains more than $10^{11}$ edges. Managing so many edges to search for a maximum weight clique is a very challenging task! So, an effective genuine exact algorithm is needed to solve MWVC exactly in large sparse graphs.

\section{Graph Reductions}

In this section, we introduce some novel reduction rules for MWVC and prove their soundness. Applying each reduction rule to a vertex-weighted graph $G$, we obtain a simplified graph $G'$. Then a minimum weight vertex cover $S'_m$ can be computed from $G'$ more easily than from $G$. For each reduction rule, we specify the relation of a minimum weight vertex cover $S_m$ of $G$ with $S'_m$, so that $S_m$ can be easily obtained from $S'_m$.

All the reductions rules are based on the following fundamental property of a vertex cover: if a vertex $v$ is not in a vertex cover, then all the neighbours of $v$ should be in the cover. As will be shown later, these rules are crucial to solve MWVC in large sparse graphs.

\subsubsection{Degree-0  Rule} For any vertex $v \in V$, if $d(v)=0$, then $G' = G \setminus \{v\}$ and $S_m = S'_m$.

\subsubsection{Adjacent  Rule} For any vertex $v \in V$, if $w(v)\ge w(N(v))$, then $G' = G \setminus N^*(v)$ and $S_m = S'_m \cup N(v)$.

\vspace{3mm}
\noindent \textit{Proof}. It is easy to see that $S'_m \cup N(v)$ is a vertex cover of $G$ and has smaller weight than $S'_m \cup \{v\}$. So, $S'_m \cup N(v)$ is a minimum weight vertex cover of $G$. $\Box$


\subsubsection{Example 1}Let $G=(V,E,w)$ be the vertex-weighted undirected graph of Figure 1, where $v_i^w$ denotes vertex $v_i$ with weight $w(v_i)$. Obviously, $w(v_5)>w(v_4)+w(v_3)$. Then we can remove $v_5$, $v_3$ and $v_4$ to obtain a subgraph $G' = ( \{v_1,v_2,v_6\}, \{(v_1, v_2)\}, w)$ by Adjacent Rule. It is easy to see $S'_m=\{v_1\}$ for $G'$. Because $N(v_5) = \{v_3, v_4\}$, $S_m=S'_m \cup N(v_5) =\{v_1,v_3,v_4\}$.    $\Box$

\begin{figure}[htbp]
\centering
\includegraphics[scale=0.34]{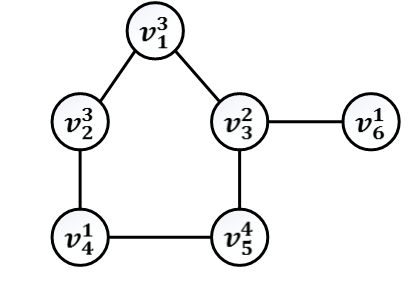}
\caption{A vertex-weighted undirected graph $G$ }\cite{jiang2017exact}
\end{figure}

\subsubsection{Degree-1  Rule} For any vertex $v \in V$, if $w(v)\leq w(N_1(v))$, then $G' = G \setminus N^*_1(v)$, and $S_m = S'_m \cup \{v\}$. 

\vspace{3mm}
\noindent \textit{Proof}. Note that vertices in $N_1(v)$ do not have any neighbour other than $v$. Either $v$ or all vertices in $N_1(v)$ should be in a vertex cover of $G$ to cover the edges between $v$ and the vertices in $N_1(v)$. Since $w(v)\leq w(N_1(v))$, $S_m = S'_m \cup \{v\}$ covers all edges of $G$ with minimum weight.

\subsubsection{Example 2}Let $G=(V,E,w)$ be the vertex-weighted undirected graph of Figure 1 but the weight of $v_6$ is 3 (instead of 1). Since $N^*_1(v_3) = \{v_3, w_6\}$ and $w(v_3) < w(v_6)$, we can remove $v_3$ and $v_6$ from $G$ to obtain $G'$. We then compute easily $S'_m=\{v_1,v_4\}$ for $G'$ and $S_m = S'_m \cup \{v_3\} = \{v_1,v_3,v_4\}$ for $G$. $\Box$





\subsubsection{Degree-2  Rule}Let $A \subseteq V$ be a set of vertices with degree 2 that are all adjacent to vertices $v_i$ and $v_j$. If $w(v_i)+w(v_j) \leq w(A)$, then $G' = G\setminus (A \cup \{v_i, v_j\})$ and $S_m = S'_m \cup \{v_i, v_j\}$.






\vspace{3mm}
\noindent \textit{Proof}. Note that there are no edges between any vertex in $A$ and any vertex other than $v_i$ and $v_j$ in $G$. It is easy to see that all edges that are in $G$ but not in $G'$ are covered by $\{v_i, v_j\}$. In addition, $w(v_i)+w(v_j) \leq w(A)$. Therefore, the set $S'_m \cup \{v_i, v_j\}$ covers all edges of $G$ and has minimum weight. 
$\Box$

\subsubsection{Example 3} Let $G=(V,E,w)$ be the vertex-weighted undirected graph of Figure 1. Consider vertices $v_3$ and $v_4$ and $A =\{v_5\}$, because $v_5$ is adjacent to both $v_3$ and $v_4$ and has degree 2. Since $w(v_3) + w(v_4) < w(A)$, we can remove $v_3$, $v_4$ and $v_5$ from $G$ to obtain $G'$. Then, $S'_m = \{v_1\}$ is a minimum weight vertex cover of $G'$ and $S_m = S'_m \cup \{v_3, v_4\} = \{v_1, v_3, v_4\}$ is a minimum weight vertex cover of $G$. $\Box$

Note that when $|N_1(v)| = 1$ or $|A| =1$, Degree-1 Rule and Degree-2 Rule are equivalent to Adjacent Rule. To see this, let $N_1(v)=\{u\}$. Then, $N(u) = \{v\}$. So, $w(v) \leq w(N_1(v))$ is equivalent to $w(u) \geq w(N(u))$. Applying Adjacent Rule to $u$ is equivalent to applying Degree-1 Rule to $v$ in this case. The rules are different when $|N_1(v)| > 1$ or $|A| >1$. For example, if $N_1(v)=\{u_1, u_2\}$, $w(v) > w(u_1)$, $w(v) > w(u_2)$, but $w(v) \leq w(u_1) + w(u_2)$, Adjacent Rule is not applicable, but Degree-1 Rule is applicable.

\subsection{Reduction Algorithm}

We implemented Degree-0  Rule, Adjacent Rule, Degree-1 Rule and Degree-2 Rule in a reduction procedure to simplify $G$. The application of each rule results in a simplified graph $G'$. Algorithm 1 shows the reduction procedure, which applies the rules in the ordering Degree-0  Rule, Adjacent  Rule, Degree-1  Rule and Degree-2  Rule, each rule being repeatedly applied until it is no longer applicable. This ordering is chosen according to the increasing complexity of the rules. 

Note that when a rule is applied after another rule, it can make the other rule applicable again. This is why the rule applications are implemented in a ``do while" loop in Algorithm 1 to simplify as much as possible the graph. For example, when Degree-2 Rule is applied to remove $v_3$, $v_4$ and $v_5$ from $G$ in Figure 1 (See Example 3), the degree of $v_6$ becomes 0, enabling Degree-0 Rule. The loop stops when no rule is applicable. After using Algorithm 1, we obtain a simplified graph and a partial vertex cover, which is of great significance to the subsequent search.

\begin{algorithm}[ht]
\caption{Reduce($G$)}\label{algorithm}
\KwIn {a graph $G=(V, E, w)$}
\KwOut{a reduced graph $G'$ and a partial vertex cover $S$
}
  \Begin{
 $S \leftarrow \emptyset$\; 
 \SetKwRepeat{doWhile}{do}{while}

    \doWhile{$|V'|\neq |V|$}{
    $V' \leftarrow V$\;
    \While {$v$ in $V$ satisfies Degree-0 Rule} {$G \leftarrow$ \textit{Degree0Rule}($G$); }
   \While {$v$ in $V$ satisfies Adjacent Rule} {$G \leftarrow$ \textit{AdjacentRule}($G$);\\ move the vertices of $N(v)$ from $V$ to $S$;}
   \While {$v$ in $V$ satisfies Degree-1 Rule} {$G \leftarrow$ \textit{Degree1Rule}($G$);\\  move $v$ from $V$ to $S$;}
   \While {$v_i$ and $v_j$ in $V$ satisfies Degree-2 Rule} {$G \leftarrow$ \textit{Degree2Rule}($G$);\\  move $v_i$ and $v_j$ from $V$ to $S$;}
    }

\Return $G[V']$ and $S$ \; 
}
\end{algorithm}

%
%
%
%

\begin{algorithm}[ht]
\caption{BMWVC$(G)$}\label{algorithm}
\KwIn{a graph $G=(V, E, w)$}
\KwOut{A minimum weight vertex cover of $G$}

($G', S$) $\leftarrow$ Reduce($G$)\;
Let $G'$ consist of disjoint subgraphs $G_1=(V_1, E_1, w)$, $G_2=(V_2, E_2, w)$, ...,$G_k=(V_k, E_k, w)$; /*$k\geq 1$*/ \\
 \Return $S$ $\cup$ Search$(G_1, \emptyset, V_1)$ 
 $\cup$ Search$(G_2,  \emptyset, V_2)$ 
 $\cup \cdots \cup$ Search$(G_k, \emptyset, V_k)$\;
\end{algorithm}

\begin{algorithm}[ht]
\caption{Search$(G, S, S_b)$}\label{algorithm}
\KwIn{a graph $G(V, E, w)$, a growing vertex cover $S$, the best vertex cover $S_b$ found so far}
\KwOut{a minimum weight vertex cover of $G$ extended from $S$}

\If{$G$ is empty} {
	\Return $S$;
	}
\If {LowerBound$(G)+w(S) \geq w(S_b)$} { 
   	\Return $S_b$;
	}	
select a vertex $v$ from $G$ using a heuristic\;
$S_b \leftarrow$ Search$(G\setminus \{v\}, S \cup \{v\}, S_b)$\;
\Return Search$(G\setminus N^*(v),  S \cup N(v), S_b)$\;
\end{algorithm}

\section{BMWVC: A Branch-and-Bound Algorithm for MWVC}

In this section, we present our branch-and-bound algorithm BMWVC for MWVC, which is shown in Algorithm 2. Given a vertex-weighted graph $G$,
 BMWVC begins by simplifying $G$ using the reduction rules. The reduction produces a set of vertices $S$ that must be included in a minimum weight vertex cover of $G$, and can result in several disjoint subgraphs of $G$: $G_1, G_2, \ldots, G_k$ ($k\geq 1$). Then, BMWVC calls a Search procedure for each of the subgraph $G_i$ to independently searching for a minimum weight vertex cover of $G_i$. The union of these minimum weight vertex covers, together with $S$ and $S'$, is returned as an optimal solution of $G$.

The Search procedure, showed in Algorithm 3, searches for a minimum weight vertex cover of $G$, beginning from an empty growing partial vertex cover $S$. It uses the best solution $S_b$ of $G$ found so far to prune search, because it should find a solution better than $S_b$ for $G$. For this purpose, it computes a lower bound of the best solution of $G$ and compares the lower bound with $S_b$. If the lower bound is smaller than the total weight of $S_b$, the search cannot be pruned and a vertex is chosen for branching. A minimum weight vertex cover of $G$ containing $\{v\}$ and another not containing $v$ (so, it should contain all vertices of $N(v)$) are respectively searched for, and the best one is returned. Note that a solution better than $S_b$ is returned only when $G$ becomes empty in a recursive call, because only in this case, the search was not pruned because of $S_b$. In all other cases, the Search procedure returns $S_b$.

There are two crucial components in the Search procedure. The first component is a branching heuristic that selects a vertex $v$ and implicitly divide the set of all vertex covers of $G$ into the subset of vertex covers that contain $v$ and the subset of vertex covers that do not contain $v$. In the first case, the search continue to search for a minimum weight vertex cover of $G\setminus \{v\}$, and in the second case, the vertices in $N(v)$ should be in the vertex covers in the subset and the search continue to search for a minimum weight vertex cover of $G\setminus N^*(v)$.

The branching heuristic is important because it may impact greatly on the search tree size. We consider the following four heuristic strategies related to degree and weight of a vertex. 

\begin{itemize}
\item \textbf{H1}: select a vertex $v$ from $V$ with the greatest $d(v)$.
\item \textbf{H2}: select a vertex $v$ from $V$ randomly.
\item \textbf{H3}: select a vertex $v$ from $V$ with the smallest $w(v)$.
\item \textbf{H4}: select a vertex $v$ from $V$ with the greatest $d(v)/$ $w(v)$.
\end{itemize}

The reason why we take degree and weight as the heuristic factors is that more vertices and edges can be deleted after a vertex with the greater $d(v)$ is selected, which can reduce the size of the explored search tree. In addition, since the aim of the MWVC problem is to find a minimum weight vertex cover, we prefer to select a vertex with a smaller weight. When there  are more than one vertex as candidates, we select a vertex $v$ with the smallest number of edges whose endpoints are both in $N(v)$.


The second  component of BMWVC is a lower bound of $S_m$, which is essential to reduce the search space. If the weight of current partial vertex cover plus the lower bound is equal to or bigger than the weight of the best solution found so far, we can prune the current subtree. The tighter the lower bound is, the more search space can be pruned. BMWVC adopts an easy and natural method to calculate the lower bound. It partitions $G$ into a set of disjoint cliques $C_1$, $C_2$, ...,$C_n$. Then \textit{LowerBound}(\textit{G}) $=\sum\limits_{i=1}^{n}(w(C_i)-\max\limits_{v_j\in C_i}(w(v_j)))$, because $k-1$ vertices are needed to cover all edges of a clique of $k$ vertices.

\begin{table}[htb]
 \scriptsize
  \centering
  \caption{Runtimes in seconds of 4 branching heuristics for a set of graphs. The cut-off time is 1000 seconds}
    \begin{tabular}{	L{2.5cm}L{1cm}L{1cm}L{1cm}L{0.9cm}L{0.9cm}L{0.9cm}L{0.4cm}}
    \toprule
\multirow{1}[0]{*}{Instance}&H1 &H2 &H3 &H4  \\
\midrule
bio-celegans	&\textbf{0.604}	&-&-&-\\
bio-diseasome	&\textbf{0.031}	&29.946	&10.322	&0.553\\
bio-yeast	&\textbf{0.005}	&0.008	&0.008	&0.007\\
bn-macaque-rhesus-brain-2	&0.001	&0.001	&0.001	&0.001\\
boyd2	&\textbf{0.703}&-&-&-\\
ca-CSphd	&0.007	&0.007	&0.007	&\textbf{0.006}\\
ca-Erdos992	&\textbf{0.006}	&\textbf{0.006}	&0.007	&0.007\\
ca-netscience	&\textbf{0.025}	&265.614	&52.346	&0.530\\
can-96	&0.104	&1.435	&1.102	&\textbf{0.019}\\
com-youtube	&\textbf{0.092}	&0.097	&0.097	&0.099\\
dwt-59	&0.251	&0.153	&0.717	&\textbf{0.137}\\
gent113 	&\textbf{0.017}	&0.290	&0.161	&\textbf{0.017}\\
GD95-c	&\textbf{0.017}	&0.073	&0.066	&0.030\\
ia-email-EU	&\textbf{0.029}	&\textbf{0.029}&\textbf{0.029}	&0.030\\
ia-enron-only	&\textbf{0.568}	&-	&-	&52.192\\
ia-reality	&\textbf{0.011}	&\textbf{0.011}& \textbf{0.011}	&0.012\\
inf-contiguous-usa	&\textbf{0.013} &0.025	&0.083	&0.026\\
lp1	&0.328	&0.344	&0.361	&\textbf{0.292}\\
opsahl-southernwomen	 &0.003	&0.002	&0.002	&\textbf{0.001}\\
road-chesapeake	&\textbf{0.003} 	&0.012	&0.026	&0.008\\
rt-retweet	&\textbf{0.002}	&0.030	&0.003	&0.003\\
rt-retweet-crawl	&\textbf{0.545}	&18.811	&19.172	&3.325\\
rt-twitter-copen	&0.005	&0.005	&\textbf{0.004}	&0.005\\
soc-dolphins	&\textbf{0.007}	&0.043	&0.029	&0.020\\
soc-douban	&0.149	&\textbf{0.122}	&0.123	&\textbf{0.122}\\
soc-gplus	&0.023	&\textbf{0.022}	&\textbf{0.022}	&0.030\\
soc-karate	&0.002	&0.002	&0.002	&\textbf{0.001}\\
tech-as-caida2007	&\textbf{0.036}	&0.068	&0.086	&0.048\\
web-edu	&0.055	&0.076	&0.108	&\textbf{0.047}\\
web-indochina-2004	&\textbf{2.572}	&-&-&-\\
web-google	&\textbf{0.031}&0.150	&0.006	&0.061\\
web-polblogs	&\textbf{0.829}&-&-&-\\
web-webbase-2001	&\textbf{0.566}	&-&-&-\\

  \bottomrule
    \end{tabular}%
  \label{tab:table1}%
\end{table}%

\section{Experiments}

In this section, we evaluate our BMWVC on real-world massive graphs from Network Data Repository \cite{nr}, which are used to evaluate MWVC and MWC problems in \cite{cai2018improving,jiang2017exact} and the graphs from practical applications \cite{mccreesh2017maximum}. The weights in the first dataset are assigned $w(v_i) = (i+1)\ mod\ 200$ to each vertex $v_i$ as in \cite{cai2018improving,jiang2017exact}. The weights in the second dataset represent real meanings. We implement our algorithm BMWVC in Java and compile it using jdk (version 1.7). The experiments are carried out on a workstation under CentOS Linux, using 12 core Intel(R) Xeon(R) E5-2650 v4  2.20 GHz CPU and 128 GB RAM. In all experiments, the runtimes are in seconds and ``-'' means timeout or out of memory.

\subsection{Heuristic Selection}

In the subsection, we test four heuristic strategies H1, H2, H3, and H4 described above to select the best one. Table 1 shows the running time of the four heuristic strategies with a cut-off time of 1000 seconds. From the table, we can find that H1 performs the best on most instances. In contrast, H2, H3 and H4 fail to solve more than 5 instances, which can be solved by H1 very easily. On the whole, H1 is the best heuristic strategy and we select it as the branching heuristic of BMWVC in all subsequent experiments.

\begin{table}[h]
\scriptsize
  \centering
  \caption{Remaining number of vertices after applying one or several reduction rules to a graph. The last column gives the number of subgraphs after applying all reduction rules in $R_{a12}$}
    \begin{tabular}{L{2.5cm}L{1cm}L{1cm}L{1cm}L{1cm}L{1cm}L{1cm}L{0.4cm}}
    \toprule
\multirow{1}[0]{*}{Instance} &$|V|$ & $R_{a}$ & $R_{a12}$  &$N_{G}$ \\
\midrule
bio-celegans     										& 453    &422		&419	 		&1\\
bio-diseasome           								& 516    &335   		&303			&13\\
bio-yeast													& 1458		&297	&118					&18	\\
bn-macaque-rhesus-brain-2          			&91    	&0	&0								&0\\
boyd2  														& 466316		&140180		&120666	&1	\\
ca-CSphd													& 1882		&394	&77				&14			\\
ca-Erdos992  											& 5094		&429	&0				&0				\\
ca-netscience         									& 379   	 &275   		&257			&22 \\
can-96           											& 96    	&96	& 96    					   	&1\\
com-youtube  											& 6320737	&653		&0				&0	\\
dwt59                                   						& 59    &51	&51							&1\\
ia-email-EU  												& 32430		&605	&0				&0								\\
ia-enron-only            									& 143	 & 138   	&138 			&1\\
ia-reality  													& 6809		&0			&0				&0												\\
inf-contiguous-usa             						& 49    	 &42		&42						&1\\
GD95-c                                 						& 62    &62	&62							&1\\
gent113            											& 113	&104		&104					&1\\
lp1  															& 534388		&5115				&2831		&320						\\
opsahl-southernwomen								& 18 		& 18   		&18			 	 	&1\\
road-chesapeake                						& 39    	&39		&39					&1 \\
rt-retweet-crawl  										& 1112702		&89603	&2309				&289							\\
rt-retweet													&96		&24 			&5					&1	\\
rt-twitter-copen	 									& 761	 &248		&65				&13\\
soc-dolphins                       						& 62    	&50			&50			&1\\
soc-douban  											& 154908		&26164 	&38				&4	\\
soc-gplus   												& 23628		&115	&0 		&0									\\
soc-karate                           						& 34 	&20			 &8					&1\\
tech-as-caida2007  									& 26475	&2657	&272			&46							\\
web-edu													& 3031		&1177	&985			&28								\\
web-indochina-2004  								& 11358	&9280	&9026			&483									\\
web-google  												& 1299		&516		&363			&24				\\
web-polblogs          									& 643    &255   		&194 	 	&1	\\
web-webbase-2001									& 16062	&3141	&1996		&76	\\

  \bottomrule
    \end{tabular}%
  \label{tab:table2}%
\end{table}%

\subsection{Effectiveness of Reduction Rules}

To evaluate the effectiveness of the proposed reduction rules, we implemented two different reduction algorithms: $R_{a}$ and $R_{a12}$. Specially, $R_{a}$ employs Adjacent reduction and $R_{a12}$ use Adjacent Rule, Degree-1 Rule and Degree-2 Rule and is exactly Algorithm 1. Note that the trivial Degree-0 rule is executed in all reduction algorithms. 

Table 2 shows the remaining vertices after performing the three reduction algorithms to simplify a set of graphs. Since the application of these reduction rules to a graph can result in several subgraphs, we also count the number of subgraphs, denoted by $N_G$ after applying $R_{a12}$. From the table, we see that reduction rules are effective. Comparing the size of the graph after the reduction, we find $R_{a12}$ produces smaller graphs than $R_{a}$. 

We also studied the effectiveness of our reduction rules. Figure 2 compares BMWVC and OBMWVC, a variant of BMWVC without using any reduction rule. It is easy to see from Figure 2 that BMWVC is much faster than OBMWVC.

\begin{figure}[htbp]
\centering
\includegraphics[scale=0.75]{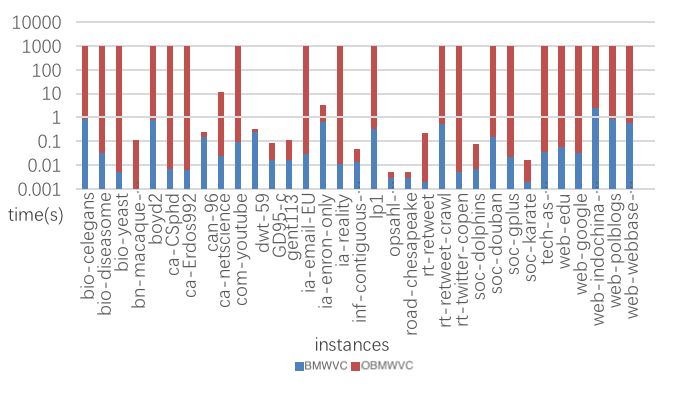}
\caption{Runtimes of BMWVC(blue) and OBMWVC(red)}
\end{figure}

\begin{table}[htbp]
\scriptsize

		\begin{center}
		\centering
		\caption{Experimental results in large spare graphs}
			\begin{tabular}{L{1.3cm}L{1cm}L{0.1cm}L{0.1cm}L{0.1cm}L{0.9cm}L{0.4cm}L{0.4cm}}
				\toprule
				
				\multicolumn{1}{l}{Instance}
				& \multicolumn{1}{l}{$S_m$}
				& \multicolumn{1}{l}{BMWVC}
				& \multicolumn{1}{l}{WLMC}
				& \multicolumn{1}{l}{SBMS}
				& \multicolumn{3}{l}{Dynwvc1/Dynwvc2}\\
				
				& & & & &best&avgt\\
				\midrule
bio-celegans     			& 20099			&\textbf{0.60}				 		&6.04		 		&- 		&20100$\uparrow$	&0.03\\
bio-diseasome           	& 22205    		&0.03   								& 377.82			&-			&22205					&\textbf{\textless 0.01}\\
bio-yeast						& 36168			&\textbf{\textless 0.01}			&- 					&- 		&36168						&\textbf{\textless 0.01}\\
bn-macaque-rhesus-brain-2          &284   		&\textbf{\textless 0.01}		&0.15			&-			&284						&\textbf{\textless 0.01} \\
boyd2  							& 17025951 		&\textbf{0.70}						&- 			&- 				&17032631$\uparrow$				&0.71\\
ca-CSphd						&48899			&\textbf{\textless 0.01}			&- 			&- 				&48899					&0.02\\
ca-dblp-2010				&15636204*		&90.41*								&-				&-					&12156739			&966.98\\
ca-dblp-2012				&19995803*		&127.29*							&-				&-					&15026140				&988.05\\
ca-Erdos992  				&40710			&\textbf{\textless 0.01}			&-				&-					&41110	$\uparrow$	&\textbf{\textless 0.01}\\
ca-netscience         		&19292   	 		&0.03  						&0.50 			&201.88				&19292 					&\textbf{\textless 0.01}	\\
can-96           				&3360 				&0.10						&0.19			&7.37				&3360						&\textbf{\textless 0.01}\\
com-youtube  				&58549			&\textbf{0.09}			&-					&-						&58549					&0.71\\
dwt59                              & 1031   			&0.25						&0.16			&3.58				&1031					&\textbf{\textless 0.01}\\
GD95-c                           & 1215 				&0.02						&0.16			&3.63				&1215					& \textbf{\textless 0.01}	\\
gent113            				& 3152				&0.02						&0.15			&12.30				&3152					&\textbf{\textless 0.01}\\
ia-email-EU  					& 78310			&\textbf{0.03}			&-					&-						&78311$\uparrow$	&\textless 0.01	\\
ia-enron-only            		& 5988	 			&0.57   					&0.27	 		&-	 					&5988						&\textbf{\textless 0.01}\\
ia-reality  						&3601				&0.01						&-					&-						&3601						&\textbf{\textless 0.01}	\\
inf-contiguous-usa        & 741    			&0.01						&0.14			&3.39				&741						&\textbf{\textless 0.01}\\
inf-roadNet-CA				&194721905*	&0.0*						&-					&-						&97206712				&999.92\\
inf-roadNet-PA				&68980405*	&986.45*				&-					&-						&53268783 				&416.69\\
lp1  								&24165495		&\textbf{0.33}			&-					&-		  				&24190047$\uparrow$		&103.32\\
opsahl-southernwomen		&97 				&\textbf{\textless 0.01}	&0.24		&3.07  				&97 					&\textbf{\textless 0.01}\\
rec-amazon 					&4071861 		&\textbf{17.56} 		&- 				&- 					&4121394$\uparrow$	 &813.93\\
road-chesapeake           &446   				&\textbf{\textless 0.01}	&0.14		&3.22				&446								&\textbf{\textless 0.01}\\
rt-retweet						&1339				&\textbf{\textless 0.01} 	&0.15 		&10.16 				&1339								&\textbf{\textless 0.01}	\\
rt-retweet-crawl  			&7821627 		&\textbf{0.55}				&-				&-						&7821858$\uparrow$		&816.47	\\
rt-twitter-copen	 		&19353	 			&\textbf{\textless 0.01}	&- 				&- 				&19353								&\textbf{\textless 0.01}\\
soc-dolphins                 &969   				&\textbf{\textless 0.01}	&0.14			&5.41			&969								&\textbf{\textless 0.01}\\
soc-douban  				&857460			&\textbf{0.15}				&-					&-					&857462$\uparrow$			&1.70\\
soc-flixster  					&9428344 		&\textbf{1.21} 				&- 				&-					&9529398$\uparrow$		&740.35\\
soc-FourSquare			&63580619*		&0.0*							&-					&-					&8768847 						&812.02\\
soc-gplus   					&12406				&\textbf{0.02}				&-					&-					&12406									&0.03		\\
soc-karate                     &226 				& \textbf{\textless 0.01}					&0.02			&3.14 &226							&\textbf{\textless 0.01}\\
soc-lastfm 					&7698963 		&\textbf{0.92} 		&- 					&- 				&7778530$\uparrow$		&471.64\\
soc-twitter-follows 		&231887 			&\textbf{0.19}    		&- 				&- 				&234249$\uparrow$	&1.06			\\
socfb-A-anon				&308165360*	&0.0*  					&-						&-					&37022624	&188.89\\
socfb-B-anon  				&29555630 		&\textbf{4.85}			&-					&-					&29901404$\uparrow$		&999.61\\
tech-as-caida2007  		& 310840			&\textbf{0.04}			&- 				&-					&310860$\uparrow$			&50.26			\\
tech-internet-as			&487337 			&\textbf{0.06}			&- 				&-	 				&492656$\uparrow$	    &61.57		\\
tech-ip							&6653664		&\textbf{1.48}			&- 				&-			 	 	 &6653664				 &720.43\\
web-edu						& 138214			&0.06					& -			&- 				&138214						&\textbf{0.03}	\\
web-indochina-2004  	&713983			&\textbf{2.57}		& -			&-					&714003$\uparrow$		&268.00	\\
web-google  					&41511				&0.03					&- 			&-					&41511	  						&\textbf{0.02}	\\
web-polblogs          		&20093    			&0.83   				&-	 			&-					&20093									&\textbf{\textless 0.01}\\
web-webbase-2001		&254341			&\textbf{0.57}		&-				&-					&254343$\uparrow$		&75.93		\\
web-wikipedia2009		&185508394*	&0.0*					&-				&-					&57518328						&330.00\\

  \bottomrule

		\end{tabular}
		 \label{tab:table3}%
		\end{center}
		\end{table}

\subsection{Comparison in Large Spare Graphs}

We conducted an experiment on real-world large graphs to compare BMWVC with WLMC \cite{jiang2017exact}, the most competitive exact solver for MWC, SBMS \cite{xu2016new}, a solver encoding MWVC into SAT, Dynwvc1 and Dynwvc2 \cite{cai2018improving}, two best heuristic MWVC solvers. The cut-off time of all solvers is 1000 seconds.  For MWC solver WLMC, we execute it for the complementary graphs. For the sake of space, we only show the results on the 47 graphs with $|E|/|V|$ less than 7. 

Table 3 indicates runtimes of the five solvers. The first column gives the name of the graphs. The second column $S_m$ gives the solutions MWVC of these graphs found by BMWVC. BMWVC proved the optimality of these solutions,  except those marked by ``*", where ``*" indicates that BMWVC encountered memory overflow and stopped the search and that the corresponding solutions are the best solutions found by BMWVC before memory overflow. The next three columns gives the runtimes of BMWVC, WLMC and SBMC. The last two columns give the best solutions found by two heuristics Dynwvc1 and Dynwvc2 over the 10 independent runs (the cut-off time of each run is 1000 seconds) and the mean time to reach the best solution in each run (avgt), ``$\uparrow$” indicating that the heuristic algorithms don't find the optimal solution. The best times are in bold (for heuristic solvers, times are not in bold if the best weight found is not optimal or has not been proved to be optimal by an exact solver). 

From table 3, we observe that BMWVC finds the optimum for most graphs, while Dynwvc1 and Dynwvc2 fail to find the optimum on 16 instances; WLMC and SBMS fail to solve more than 32 instances. Surprisingly, BMWVC needs less time than heuristic solvers on 17 instances to find and prove the optimal solution. In general, these results indicate that BMWVC is an extremely competitive exact solver.

\subsection{Comparison in Application Graphs}

We also compare our BMWVC with MLMC, SBMS, Dynwvc1, and Dynwvc2 in application graphs, whose densities (($2*|E|)/(|V|*(|V|-1))$) range from 0.04 to 0.98. These graphs, which are used to evaluate MWC solvers in \cite{jiang2018two,hebrard2018conflict}, are derived from four important practical problems: the winner determination problem (WDP), error-correcting codes (ECC), kidney-exchange schemes (KES), and the research excellence framework (REF).  Table 4 indicates the number of successfully solved instances grouped by families within a cut-off time of 5 hours for exact solvers and heuristic solvers, where ``$\#$'' is the total instances of each family. Note that the MWVC solvers BMWVC, SBMS, Dynwvc1, and Dynwvc2 are performed on the complementary graphs. 

From Table 4, it is easy to see that BMWVC performs significantly better than SBMS, Dynwvc1 and Dynwvc2. It performs slightly less well than WLMC,  because these graphs arise from MWC practical applications, and some of them are too sparse so that the complementary graphs are too dense for all MWVC solvers. The purpose of this experiment is to show that  BMWVC is competitive even for these applications not exactly in its field and is substantially better than other MWVC solving approaches.  

\begin{table}[htbp]
 \scriptsize
  \centering
  \caption{The number of solved graphs }
    \begin{tabular}{L{1cm}L{1cm}L{1cm}L{1cm}L{1cm}L{1cm}L{1cm}L{1cm}L{0.4cm}}
    \toprule
\multirow{1}[0]{*}{Instance} &$\#$ & BMWVC & WLMC & SBMS & Dyn1/Dyn2   \\
\midrule
WDP &50			&50		&50		&10		&39 \\
KES	&100			&89		&100		&71		&27\\
ECC	&15			&13		&15		&1		&15\\
REF	&128			&103		&104		&4		&128\\
total   &293		&255	&269	&86		&209 \\                   
  \bottomrule
    \end{tabular}%
  \label{tab:table4}%
\end{table}%

\section{Conclusions}

We proposed BMWVC, a new exact MWVC algorithm, which involves four reduction rules, a tight lower bound, and an efficient branching heuristic. The reported experiments show that the reduction rules are very effective in reducing massive graphs to tractable sizes; the tight lower bound can successfully prune search; and the efficient branching heuristic greatly decreases the running time. They also show that BMWVC outperforms relevant exact and heuristic MWVC solving approaches for real-world large graphs and application instances.

\bibliographystyle{named}
\bibliography{ijcai19}

\end{document}